\documentstyle[preprint,eqsecnum,aps]{revtex}

\begin{document}

\title{Decoherence in Phase Space}
\author{{{M. Murakami$\dagger$, G.W. Ford$\ddagger$, and R.F. O'Connell$\dagger$} 
\\
{\it{$\dagger$Department of Physics and Astronomy, Louisiana State University, \\ 
Baton Rouge, LA 
70803-4001} \\
$\ddagger$Department of Physics, University of Michigan, \\ Ann Arbor, MI  48109-
1120}}}

\date{\today}

\maketitle

\begin{abstract}
Much of the discussion of decoherence has been in terms of a particle moving in 
one
dimension that is placed in an initial superposition state (a Schr\"{o}dinger 
"cat"
state) corresponding to two widely separated wave packets.  Decoherence refers to 
the
destruction of the interference term in the quantum probability function.  Here, 
we
stress that a quantitative measure of decoherence depends not only on the specific
system being studied but also on whether one is considering coordinate, momentum 
or phase space.  We show
that this is best illustrated by considering Wigner phase space where the measure 
is again different. 
Analytic results for the time development of the Wigner distribution function for 
a two-Gaussian
Schr\"{o}dinger "cat" state have been obtained in the high-temperature limit 
(where decoherence can occur
even for negligible dissipation) which facilitates a simple demonstration of our 
remarks. 
\end{abstract}


\newpage

\section{Introduction}

Decoherence refers to the destruction of a quantum interference pattern and is 
relevant to
the many experiments that depend on achieving and maintaining entangled states. 
Examples of such efforts are in the areas of quantum
teleportation\cite{zeilinger001}, quantum information and
computation\cite{bennett95,quantuminfo98}, entangled states\cite{haroche98},
Schr\"{o}dinger cats\cite{zeilinger002}, and the quantum-classical
interface\cite{tegmark01}.  For an overview of many of the interesting experiments
involving decoherence, we refer to Refs. \cite{haroche98} and \cite{myatt00}.

Much of the discussion of 
decoherence\cite{giulini96,paz01,ford011,ford012,ford013} has
been in terms of a particle moving in one dimension that is placed in an initial
superposition state (a Schr\"{o}dinger "cat" state) corresponding to two widely 
separated
wave packets, each of the same form but having their centers $x_{0}$ at $x_{0}=\pm 
d/2$ so that the packets
are separated by a distance $d$.  Thus, in an obvious notation we write the wave 
function of
the superposition state as

\begin{equation}
\psi (x,t)=N[\psi_{1}(x,t)+\psi_{2}(x,t)], \label{dps1}
\end{equation}
where $\psi_{1}$ and $\psi_{2}$ denote the packets with centers at $d/2$ and $-
d/2$, respectively, and $N$ is
the normalization constant.  Hence

\begin{equation}
P(x,t)=N^{2}(\vert\psi_{1}\vert^{2}+\vert\psi_{2}\vert^{2}+2{\textnormal{Re}}
\{\psi^{*}_{1}\psi_{2}\}). \label{dps2}
\end{equation}
Thus, the probability distribution consists of three contributions, two of which
correspond to the separate packets, whereas the third is an interference term.

Many investigators have considered free particle Gaussian wave packets and this 
has also
been our choice.  However, in contrast to widespread current opinion, we showed 
that it
is possible to obtain "Decoherence without Dissipation" which is actually the 
title of a
paper\cite{ford012} in which we showed that, working solely within the framework 
of
elementary quantum mechanics and equilibrium statistical mechanics, decoherence 
can in
fact occur at high temperature $T$ even for vanishingly small dissipation.  More
precisely, we consider an ensemble of particles in thermal equilibrium, but so 
weakly
coupled to a heat bath that we can neglect dissipation in the equation of motion 
so that
we have a Maxwell distribution of initial velocities\cite{ford013}.  The results 
obtained
from such a calculation are in agreement with those obtained in the appropriate 
limit
from more sophisticated calculations within the framework of nonequilibrium 
statistical
mechanics\cite{ford011,ford013}.

It is generally known that a quantitative measure of decoherence depends not only 
on the
specific system being studied but also on whether one is considering coordinate or
momentum space.  We show that this is best illustrated by considering Wigner phase 
space
where the measure is again different.  Thus, using the techniques developed in
\cite{ford012}, we obtain analytic results for the time development of the Wigner
distribution function for a two-Gaussian Schr\"{o}dinger "cat" state in the
high-temperature limit which facilitates a simple demonstration of our remarks.

As a preliminary, we consider in Sec. II the case of a free particle Gaussian wave
packet.  Starting with the position wave function $\psi (x,t)$, we then calculate 
the
corresponding momentum wave function $\tilde{\psi}(p,t)$ and the corresponding 
Wigner
distribution.  We then examine the effect of temperature on the various 
quantities.  In
Sec. III, we generalize to the two-Gaussian superposition state.  These results 
enable us
to obtain the rate of decay of decoherence in position, momentum and phase space, 
which
we discuss in Sec. IV.

\section{Single Gaussian Wave Packet}

The solution of the free-particle Schr\"{o}dinger at time $t$, given that the 
solution at
$t=0$ is a minimum uncertainty wave packet, centered at $x_{0}$ and moving with 
velocity $v_{0}$, is
\cite{ford012}

\begin{equation}
\psi (x,t) = \frac{1}{\left[2\pi\left(\sigma
+\frac{i\hbar
t}{2m\sigma}\right)^{2}\right]^{1/4}}\exp\left\{-\frac{(x-x_{0}-
v_{0}t)^{2}}{4\sigma^{2}+(2i\hbar
t/m)}+i\frac{mv_{0}}{\hbar}x-i\frac{mv^{2}_{0}t}{2\hbar}\right\}. \label{dps3}
\end{equation}
The probability distribution is

\begin{equation}
P(x;t)=\vert\psi
(x,t)\vert^{2}=[2\pi\sigma^{2}(t)]^{-1/2}\times\exp\left\{-\frac{(x-x_{0}-
v_{0}t)^{2}}{2\sigma^{2}(t)}\right\},
\label{dps4}
\end{equation}
which is a Gaussian centered at the mean position of the particle at time $t$ with 
variance given by
$\sigma^{2}(t)=\sigma^{2}+(\hbar t/2m\sigma)^{2}$.  Without loss of generality, 
from henceforth we take
$x_{0}=0$.  It is convenient for later analysis in the two-Gaussian case to 
express these results in a more
compact form, that is

\begin{equation}
\psi
(x,t)=\left(2\pi\Sigma^{2}\right)^{-1/4}\exp\left\{i(mv^{2}/2\hbar
)t\right\}\exp\left\{-\frac{x^{2}_{1}}{4\sigma\Sigma}+i\frac{mv}{\hbar} 
x_{1}\right\}, \label{dps5}
\end{equation}
and

\begin{equation}
P(x,t)=\left[2\pi\sigma^{2}(t)\right]^{-1/2}\exp\left\{-
\frac{x^{2}_{1}}{2\sigma^{2}(t)}\right\},
\label{dps6}
\end{equation}
where

\begin{equation}
x_{1}=x-vt \label{dps7}
\end{equation}
and

\begin{eqnarray}
\Sigma &=& \sigma +i\left(\frac{\hbar t}{2m\sigma}\right) \nonumber \\
&=& \sigma +iv_{q}t. \label{dps8}
\end{eqnarray}
where

\begin{equation}
v_{q}=\frac{\hbar}{2m\sigma}. \label{dps9}
\end{equation}
Also

\begin{eqnarray}
\Sigma\Sigma^{*} &=& \sigma^{2}+\left(\frac{\hbar 
t}{2m\sigma}\right)^{2}=\sigma^{2}+(v_{q}t)^{2}
\nonumber \\
&\equiv& \sigma^{2}(t). \label{dps10}
\end{eqnarray}

The corresponding momentum wave function is

\begin{eqnarray}
\tilde{\psi}(p,t) &=& \frac{1}{(2\pi\hbar)^{1/2}}\int^{\infty}_{-
\infty}dx~\psi{(}x,t)\exp\left(
-\frac{ipx}{\hbar}\right) \nonumber \\
&=&\left(\frac{2\sigma^{2}}{\pi\hbar^{2}}\right)^{1/4}\exp\{-
i\frac{(p^{2}+m^{2}v^{2})t}{2m\hbar}
\}\exp\{-\frac{\sigma^{2}(p-mv)^{2}}{\hbar^{2}}\}, \label{dps11}
\end{eqnarray}
and hence the momentum probability distribution is

\begin{eqnarray}
P(p,t) &=& |\tilde{\psi}(p,t)|^{2} \nonumber \\
&=&
\left(\frac{2\sigma^{2}}{\pi\hbar^{2}}\right)^{1/2}\exp\{-\frac{2\sigma^{2}(p-
mv)^{2}}{\hbar^{2}}
\}. \label{dps12}
\end{eqnarray}
We note from (\ref{dps6}) and (\ref{dps10}) that the variance in coordinate space 
increases with
increasing $t$ whereas we see from (\ref{dps12}) that the variance in momentum 
space is time independent.

Going beyond our previous investigations \cite{ford012}, we now turn to the 
determination of the Wigner
distribution function $W(x,p,t)$ given by

\begin{eqnarray}
W(x,p,t)
&=&
\frac{1}{2\pi\hbar}\int_{-
\infty}^{\infty}e^{ipy/\hbar}\psi^{*}\left(x+\frac{y}{2},t\right)\psi\left(
x-\frac{y}{2},t\right)dy \nonumber \\
\nonumber \\
&=& (\pi\hbar)^{-1}\exp\left\{-\frac{X^{2}}{2\sigma^{2}}-\frac{2\sigma^{2}P^{2}}
{\hbar^{2}}\right\}, \label{dps13}
\end{eqnarray}
where

\begin{equation}
X\equiv x-\frac{pt}{m}, \label{dps14}
\end{equation}
and

\begin{equation}
P\equiv p-mv. \label{dps15}
\end{equation}

Next we consider the case of a particle in thermal equilibrium, but so weakly 
coupled to the environment
that we can neglect dissipation.  The principles of statistical mechanics then 
tell us that we obtain the
corresponding probability distribution by averaging the distribution (\ref{dps6}) 
over a thermal
distribution of velocities.  The result is

\begin{eqnarray}
P_{T}(x,t) &=&
\sqrt{\frac{m}{2\pi{k}T}}\int^{\infty}_{-\infty}dv\exp\{-
\frac{mv^{2}}{2kT}\}P(x,t) \nonumber \\
&=&
(2\pi{w}^{2})^{-1/2}\exp\left(-\frac{x^{2}}{2w^{2}}\right),
\label{dps16}
\end{eqnarray}
where

\begin{eqnarray}
w^{2}(t) &=& \sigma^{2}(t)+\frac{kT}{m}t^{2} \nonumber \\
&=& \sigma^{2}+v^{2}_{q}t^{2}+\bar{v}^{2}t^{2}, \label{dps17}
\end{eqnarray}
and

\begin{equation}
\bar{v}=\sqrt{\frac{kT}{m}}. \label{dps18}
\end{equation}
The corresponding result for the thermally averaged Wigner distribution is

\begin{equation}
W_{T}(x,p,t)=\left[(\pi\hbar)^{-
1}\left(\frac{v^{2}_{q}}{v^{2}_{q}+\bar{v}^{2}}\right)^{1/2}\right]\exp\left\{
-\frac{x^{2}}{2\sigma^{2}}-
\frac{w^{2}p^{2}}{2m^{2}\sigma^{2}(v^{2}_{q}+\bar{v}^{2})}+\frac{xpt}{m\sigma^{2}}
\right\}.
\label{dps19}
\end{equation}
As a check, we note that integration of (\ref{dps19}) over $p$ gives 
(\ref{dps16}).  Furthermore,
integration over $x$ gives

\begin{eqnarray}
P_{T}(p,t) =\left[2\pi
m^{2}(v^{2}_{q}+\bar{v}^{2})\right]^{-1/2}\exp\left\{-
\frac{p^{2}}{2m^{2}(v^{2}_{q}+\bar{v}^{2})}\right\},
\label{dps20}
\end{eqnarray}
for the momentum thermal distribution.

\section{Two-Gaussian Wave Packet}

The result has the same form as (\ref{dps1}).  For the Gaussian case, we now write 
it in the form

\begin{eqnarray}
\psi^{(2)}(x,t) &=&
N\left\{\psi\left(x-\frac{d}{2}\right)+\psi\left(x+\frac{d}{2}\right)\right\}
\nonumber \\
&=& N\exp\left\{i(mv^{2}/2\hbar
)t\right\}\exp\left(i\frac{mv}{\hbar}x_{1}\right)\left\{(2\pi\Sigma^{2})^{-
1/4}\exp
\left(-\frac{\left(x_{1}-
\frac{d}{2}\right)^{2}}{4\sigma\Sigma}\right)+(d\rightarrow
-d)\right\}, \label{dps21}
\end{eqnarray}
where

\begin{equation}
N=\left[2(1+e^{-d^{2}/8\sigma^{2}})\right]^{-1/2}. \label{dps22}
\end{equation}
It follows that \cite{ford012}

\begin{eqnarray}
P^{(2)}(x,t) &=&
N^{2}[2\pi\sigma^{2}(t)]^{-1/2}\{\exp\left(-\frac{(x_{1}-
\frac{d}{2})^{2}}{2\sigma^{2}(t)}
\right)+(d\rightarrow -d) \nonumber \\
&&
~~~~~~{}+2\exp\left(-
\frac{x^{2}_{1}+\frac{d^{2}}{4}}{2\sigma^{2}(t)}\right)\cos\frac{\hbar
tdx_{1}}{4m\sigma^{2}\sigma^{2}(t)}\}, \label{dps23}
\end{eqnarray}
and

\begin{eqnarray}
P^{(2)}_{T}(x,t) &=& N^{2}[2\pi w^{2}(t)]^{-1/2}\{\exp\left(-\frac{(x-
\frac{d}{2})^{2}}
{2w^{2}}\right)+(d\rightarrow -d) \nonumber \\
&& ~~~~~~{}+
2\exp\left(-\frac{x^{2}}{2w^{2}}-
\frac{\sigma^{2}w^{2}+(\bar{v}t)^{2}(v_{q}t)^{2}}{\sigma^{2}\sigma^{2}(t)w^{2}}~
\frac{d^{2}}{8}\right)\cos\frac{\hbar tdx}{4m\sigma^{2}w^{2}}\}. \label{dps24}
\end{eqnarray}
In addition, after some algebra, we find that the corresponding Wigner 
distribution is

\begin{equation}
W^{(2)}(x,p,t)=N^{2}\left\{W(X-
\frac{d}{2},P)+W(X+\frac{d}{2},P)+2\cos\left(\frac{Pd}{\hbar}\right)
W(X,P)\right\}, \label{dps25}
\end{equation}
where $X$ and $P$ are given in (\ref{dps14}) and (\ref{dps15}) and $W(x,p,t)$ is 
given in (\ref{dps13}).  
Also, after thermal averaging, we obtain

\begin{eqnarray}
W^{(2)}_{T}(x,p,t)&&=N^{2}\left(W_{T}(X-
\frac{d}{2},P)+W_{T}(X+\frac{d}{2},P)\right)
\nonumber \\
&&+N^{2}\left(
W_{T}(x,p,t)\exp\left\{-\frac{d^{2}} 
{8\sigma^{2}}~\frac{\bar{v}^{2}}{\bar{v}^{2}+v^{2}_{q}}
\right\}\cos\left\{\frac{pd}{\hbar}~\frac{v^{2}_{q}}{\bar{v}^{2}+v^{2}_{q}}
\right\}\right), \label{dps26}
\end{eqnarray}
where $W_{T}(x,p,t)$ is given in (\ref{dps19}).  In order to obtain the momentum 
distribution, we simply
integrate (\ref{dps25}) over $x$ to obtain

\begin{equation}
P^{(2)}(p,t)=2N^{2}P(p,t)\left\{1+\cos\left(\frac{Pd}{2\hbar}\right)\right\}, 
\label{dps31}
\end{equation}
where $P(p,t)$ is given by (\ref{dps12}).  Thus, as with $P(p,t)$, we note that 
$P^{(2)}(p,t)$ is time
independent.  Next, carrying out the thermal average, we obtain

\begin{equation}
P^{(2)}_{T}(p,t)=P_{T}(p,t)\left\{1+\exp\left(-
\frac{m^{2}d^{2}\bar{v}^{2}v^{2}_{q}}{2(\bar{v}^{2}+
v^{2}_{q})}\right)\cos\left(\frac{pd}{2\hbar}~\frac{\bar{v}^{2}}{\bar{v}^{2}+v^{2}
_{q}}\right)\right\},
\label{dps32}
\end{equation}
which, of course, is also independent of $t$.

We now have all the results necessary to discuss decoherence decay
rates, which will be the subject of Sec. 4.

\section{Decoherence Decay Rates}

In general, it is clear from the above that, in the case of a two-Gaussian 
superposition state, the
probability distribution consists of three contributions, two of which correspond 
to the separate packets,
whereas the third is an interference term.  The interference term is characterized 
by the cosine factor. 
One measures the disappearance of the interference term, that is, the loss of 
coherence (decoherence), by
defining an attenuation coefficient $a(t)$, which is the ratio of the factor 
multiplying the cosine to twice
the geometric mean of the first two terms.

Thus, in the case of decoherence in {\underline{coordinate space}}, one sees from 
an examination of
(\ref{dps24}) that

\begin{equation}
a(t)=\exp\left\{-
\frac{\frac{kT}{m}t^{2}d^{2}}{8\sigma^{4}+8\sigma^{2}\frac{kT}{m}t^{2}+\frac{2\hbar^{2}t^{2}}
{m^{2}}}\right\}. \label{dps27}
\end{equation}
For short times (characteristic of decoherence time scales), whereas the $t$ 
dependent terms in the
denominator are negligible, the $t$ dependent terms in the numerator remain, and 
thus we obtain

\begin{equation}
a(t)\cong e^{-t^{2}/\tau^{2}_{d}}, \label{dps28}
\end{equation}
where the decoherence time is

\begin{equation}
\tau_{d}=\frac{\sqrt{8}\sigma^{2}}{\bar{v}d}, \label{dps29}
\end{equation}
and $\bar{v}=\sqrt{kT/m}$ is the mean thermal velocity.  This is consistent with 
the results obtained in
Refs. \cite{ford011,ford012,ford013}, where we found that the dominant 
contribution to decoherence at high
temperatures $(kT\ll\hbar\gamma$, where $\gamma$ is typical dissipative decay 
rate), is independent of
dissipation.

Turning now to {\underline{momentum space}}, it is clear that the right-side of 
(\ref{dps32}) is independent
of time $t$.  Hence, there is no decoherence in momentum space, which is what we 
expect from physical
considerations.

Decoherence in phase space is obtained from (\ref{dps26}) leading to

\begin{equation}
a(t)=\exp\left\{\frac{d^{2}}{8\sigma^{2}}~\frac{v^{2}_{q}}{\bar{v}^{2}+v^{2}_{q}}
\right\}. \label{dps30}
\end{equation}
Thus, similar to the case with momentum space, there is no decoherence in 
{\underline{phase space}}.  We
conclude that, for the two-Gaussian superposition state, decoherence is manifest 
only in coordinate space.

\end{document}